\RequirePackage{fix-cm}
\documentclass[smallextended]{svjour3}       
\smartqed  
\usepackage{graphicx}
\usepackage{dcolumn}
\usepackage{bm}
\usepackage{amssymb}
\usepackage{amsmath}
\usepackage{amsfonts,mathrsfs}

\begin{document}

\title{Can a Bohmian be a Rovellian for all practical purposes? }


\author{Aur\'elien Drezet}


\institute{A. Drezet \at Institut NEEL, CNRS and Universit\'e Grenoble Alpes, F-38000 Grenoble, France \\
            \email{adrezet@neel.cnrs.fr} }

\date{Received: date / Accepted: date}

\maketitle

\begin{abstract}
The aim of this article is to discuss the preferred basis problem in relational quantum mechanics (RQM). The issue is at the heart of quantum mechanics  and we first show that the mathematical formalism of RQM is immune to  recent critics concerning consistency.   Moreover,  we also analyse the notion of interaction in RQM and provide a For All  Practical  Purposes (FAPP) reading of RQM comparing it with Bohmian mechanics.  
\end{abstract}

\section{Introduction}\label{sec1}
\indent  The Relational interpretation of Quantum Mechanics (RQM)~\cite{Rovelli1996,RovelliBook2021,RovelliFP} can be seen as a modern and ambitious attempt to generalize the orthodox quantum interpretation of Bohr~\cite{Bohr} and Heisenberg: The so-called Copenhagen interpretation. Like the Copenhagen interpretation RQM emphasizes the role of experimental contexts and relations between quantum systems and observers. However, unlike the Copenhagen interpretation RQM doesn't confine observers to the classical macroscopic domain (i.e., supposedly separated from the quantum realm by a shifty split or cut): RQM presupposes that quantum mechanics is universally valid in the sense that any physical system can plays the role of an observer. In contrast to what was claimed by A.~Peres~\footnote{`The two electrons in the ground state of the helium atom are correlated, but no one would say that each electron ``measures'' its partner.'~\cite{Peres}} here a single photon can be an observer for a single molecule and reciprocally. There is thus a perfect symmetry between actors and spectators, i.e., between observed and observing systems in RQM.\\ 
\indent Clearly, RQM is controversial like every quantum interpretations are. One of the most controversial issue in the RQM concerns perhaps the status of the wave function and it is the question that we will discuss in this short article. We first remind that the development of RQM is strongly motivated by the need to solve or dissolve famous quantum paradoxes involving Schr\"odinger's cat and Wigner's friends. This is indeed a prerequisite if one wants to apply quantum mechanics to gravitational and cosmological problems (the same motivation was given by Everett for developing his many worlds interpretation). Importantly, all these thought experiments proposed by Schr\"odinger, Wigner and others were originally considered as mere speculations concerning the application of quantum mechanics to large macroscopic systems involving many degrees of freedom and where decoherence plays a fundamental role (even if decoherence was not yet known by the founding fathers of quantum physics). Indeed, for large systems the classical quantum boundary becomes shifty and elusive and therefore for all practical purposes it is irrelevant for the observer to know where to locate this Heisenberg or von Neumann cut. However, with the recent development of quantum technologies it becomes possible to realize experiments with atoms, molecules, and photons mimicking some aspects of the original thought experiments. In this context the answer proposed by the Copenhagen interpretation for solving the measurement problem appears necessarily limited and approximate. The mere idea of a collapse, both epistemic and ontic, conflicts with the basic unitarity and linearity of the quantum evolution and this leads to the infamous quantum measurement problem. As emphasized by J.~S.~Bell~\cite{Bell} the vagueness of such words like measurement, observer, system, apparatus...should have no place in a `serious' physical and self-consistent theory requiring no artificial division between the small and the large, the quantum and the classical.  Indeed, if we assume the universal validity of the Schr\"odinger equation it becomes clear that soon or later with the progress of technologies we will have to experimentally fight with the shifty split introduced by the founding fathers of quantum mechanics (this is of course also true if we assume a physical spontaneaous collapse breaking unitarity).\\
\indent It is in this context that RQM appears as an interesting alternative since it assumes the universal validity of unitarity (like in the many worlds approach) but in the same time accepts the relational or perspectival methodology proposed in the Copenhagen interpretation (but without artificially imposing the macroscopic nature of the measurement procedure). This originality of RQM comes with a price since as emphasized  by Rovelli RQM realism is weakened by relationalism and facts become relative to physical systems. More precisely, as emphasized recently by M.~Bitbol~\cite{Bitbol}, in RQM the old ontology of Newton and Laplace admitting a `God view' or a `meta-context' for describing the Universe in a univocal way is abandoned and a contextual view is favored replacing `strong'  realism~\cite{Laudisa}. Clearly, this relational Leibnizian view contrasts with standard habits of classical physics and therefore RQM apparently conflicts with other approaches like the de Broglie-Bohm  (advocated by the present author) or spontaneous collapse theories assuming a classical-like basic ontology in space-time or in the configuration space.\\ 
 \indent However, it is not here the aim to give a technical review of RQM. No more than it is to provide a deep philosophical discussion concerning the status of realism or antirealism in the RQM (the author is not a philosopher).  Furthermore, since RQM is presented as being `metaphysically neutral' but actually favors a weakening of the `strong' realism of classical physics~\cite{Laudisa} it could appear as impossible for a strong `Bohmian' realist like the author to write objectively about RQM.  Moreover my interest in RQM is actually related to the role and status of the wavefunction in Rovelli's interpretation compared to what is occurring in the de Broglie Bohm mechanics. More precisely, it is my opinion that the metaphysical neutrality of RQM allows a Bohmian to find agreement with some aspects of Rovelli's framework. It was already  stressed by Bohm that Bohr's complementarity doesn't contradict the spirit of the pilot-wave theory. This is so because the pilot-wave theory relies explicitly on the quantum formalism developed by Heisenberg and Schr\"{o}dinger and because fundamental experimental limitations associated with the uncertainty principle are accepted by Bohmians. The main advantage of RQM over the Copenhagen interpretation concerns the universality of the approach consisting into separating an isolated system into an observed part (S) and an observer  (O). This democratized Copenhagen view we believe could be used as an operational formalism even by those like me  assuming an underlying `hidden' reality \`{a} la de Broglie Bohm.\\
 \indent  The aim of the present note is thus double. First, (and this is the main goal) we want to stress the metaphysical neutrality of RQM in its mathematical formulation. This actually requires to discuss shortly the role of the wave function of composite systems in RQM. In this context RQM has been recently criticized by C. Brukner~\cite{Brukner} (see also J. Pienaar~\cite{Pienaar}) and these studies actually  focus of the role of the wave function and the so called `preferred basis' ambiguity associated with the measurement problem.  Recently, Di Biagio and Rovelli offerred a detailed reply~\cite{Rovellian}. Here, the aim is not to repeat such a point by point analysis.  My own analysis of this important issue is motivated by the fact that the objections made in \cite{Brukner,Pienaar} are quite natural.  I actually did independent similar objections to myself (i.e., ignoring the analysis~\cite{Brukner,Pienaar}) and therefore the answer I found to debunk or clarify my own objections are I believe useful in this discussion \footnote{The present analysis is motivated by the recent discussion made by Rovelli at the online `quantum foundation' conference: {https://www.youtube.com/watch?v=QtU10moL-MI}, and by private  discussions with him concerning RQM and his recent book~\cite{RovelliBook2021}.  }. For this reason my reasoning will not be presented as a reply to~\cite{Brukner,Pienaar}.   This specially important since they criticized RQM using a formalism which is foreign to RQM, i.e., based on the notion of wave function for the composite system (S)+(O). In my opinion the only way to analyze RQM is to watch it from the inside and thus to see  if we agree or can complete it. In the end (and this is my second goal) I would like to briefly discuss the difference between the 	absence of a preferred basis problem in the mathematical formulation of RQM and the notion of preferred variables or beables introduced in strong realist interpretations like Bohmian mechanics. I believe that this is a confusion that could play a role in the debate and therefore I will try to clarify it a bit.   More precisely, I will argue that once the preferred basis ambiguity is formally dissolved in RQM we are still completely free to use its metaphysical (operational) neutrality to agree with the strong realist framework \`{a} la Bohm. This I believe, can be done for all practical purposes (FAPP). Moreover,  I will stress the  relation of this problem with the transition  from potential to actual advocated by Heisenberg~\cite{Heisenberg}.  I believe that this issue (related to the notion of interaction or measurement) is central to understand or perhaps clarify RQM. \\
 
\section{The preferred Basis problem in RQM} 
\indent  The main issue concerns the interpretation of the full wavefunction  $|\Psi_{SO}\rangle$ involving observer (O) and observed system (S). 
Consider for example a two-particle system where each particle (S) and (O) is represented by their spatial coordinates $x_S$ and $x_O$ respectively.  The wave function associated with the generally entangled system  reads
\begin{eqnarray}
|\Psi_{SO}\rangle=\iint dx_S dx_O\psi(x_S,x_O,t)|x_S\rangle\otimes|x_O\rangle.
\end{eqnarray} Of course, one could consider other variables for describing the (O) subsystem, like the momentum $p_O$, and we will thus obtain an alternative wave function representation \begin{eqnarray}
|\Psi_{SO}\rangle=\iint dx_S dp_O\psi'(x_S,p_O,t)|x_S\rangle\otimes|p_O\rangle.
\end{eqnarray}       
Now, suppose that we would like to built a relational quantum theory  where we introduce a distinction between the (S) and (O) subsytems and  such that we consider the description of (S) from the point of view' of (0)\footnote{We don't have to suppose that an electron has a mind or consciousness. It is for this reason that I donc accept the Brukner analysis relying on mind states in different basis.  For the problem with this concept see the note \cite{Simon} where it is shown that mind states associated with states of knowledge $|O_1\rangle$ and $|O_1\rangle$ have to be orthogonal if the observed subsystem states are: $\langle S_1|S_2\rangle=0$. In my opinion this conflicts with the postulate `DisRS' of \cite{Brukner}. }. In this relational approach we are trying to apply the method of the Copenhagen interpretation and therefore when we say that (O) is an observer for (S) we mean that the physical variables characterizing (O) are fixed or actualized as it will be in classical physics (we will come back to this problem of the actualization later).    A way to do that would be to select a preferred basis in (O) but this would somehow imply that we select a specific representation like $\psi(x_S,x_O,t)$  or $\psi'(x_S,p_O,t)$ and an additional physical postulate would have to be done.\\
\indent RQM suggests a different strategy: If we are only interested in local observables $\hat{A}_S$ of the (S) subsystem can we find a mathematical description of quantum mechanics which is independent of the  basis choice in (O)? The response is of course yes and the answer is well known: It is the reduced density matrix  
\begin{eqnarray}
\hat{\rho}^{(red.)}_{S|O}=\textrm{Tr}_O[\hat{\rho}_{SO}]=\textrm{Tr}_O[|\Psi_{SO}\rangle \langle \Psi_{SO}|]\nonumber\\
=\iiint dx_Sdx'_Sdx_0 \psi^\ast(x_S,x_O,t)\psi(x'_S,x_O,t)|x'_S\rangle\langle x_S|\nonumber\\
=\iiint dx_Sdx'_Sdp_0 \psi'^\ast(x_S,p_O,t)\psi'(x'_S,p_O,t)|x'_S\rangle\langle x_S|=...
\end{eqnarray}
With such an object we can define the mean value $\langle \hat{A}_S\rangle$ as 
\begin{eqnarray}
\langle \hat{A}_S\rangle=\textrm{Tr}_{SO}[\hat{A}_S\hat{\rho}_{SO}]=\textrm{Tr}_{S}[\hat{A}_S\hat{\rho}^{(red.)}_{S|O}]\label{mean}
\end{eqnarray}
which is clearly independent of the description used for (O). Therefore, in RQM the fundamental object  \underline{relatively to (O)} is not   $|\Psi_{SO}\rangle$ or $\hat{\rho}_{SO}=|\Psi_{SO}\rangle \langle \Psi_{SO}|$ but the reduced density matrix $\hat{\rho}^{(red.)}_{S|O}$ which is independent of the basis chosen to represent (O).  Of course, in the special case  where we have factorized states like $|\Psi_{SO}\rangle=|\Psi_{S}\rangle\otimes|\Psi_{O}\rangle$ this description simplifies and we have $\hat{\rho}^{(red.)}_{S|O}=|\Psi_{S}\rangle \langle \Psi_{S}|:=\hat{\rho}_{S}$. The independent subsystem (S) is thus a pure state described by a wave function $|\Psi_{S}\rangle $. Now, this approach opens several questions:\\ 
\indent First, what about observables $\hat{A}_{SO}$ associated with the composite system?    In that case we can not use the reduced density matrix $\hat{\rho}^{(red.)}_{S|O}$.  The solution is simple we must increase the size of the whole system and introduce a second observer (O'). Only for such an additional subsystem can we speak of a reduced density matrix for the (SO) subsystem: $\hat{\rho}^{(red.)}_{SO|O'}=\textrm{Tr}_{O'}[|\Psi_{SOO'}\rangle \langle \Psi_{SOO'}|]$.  And again, if the subsystem (O') factorizes, i.e., $|\Psi_{SOO'}\rangle=|\Psi_{SO}\rangle\otimes|\Psi_{O'}\rangle$  we have $\hat{\rho}^{(red.)}_{SO|O'}=|\Psi_{SO}\rangle \langle \Psi_{SO}|:=\hat{\rho}_{SO}$. \\
\indent This opens a different question or remark:   What about an isolated system?   Can RQM apply to it?     The answer is No. RQM consider a relational picture and we need to apply a division between two subsystems.    If we are only  one system (S')  then we must be able to split it as (S)+(O) in order to apply RQM and only then can we use this formalism by considering the two perspectives involving density matrices $\hat{\rho}^{(red.)}_{S|O}$ and $\hat{\rho}^{(red.)}_{O|S}$. In the same vein the notion of a wave function for the whole Universe makes no sense. Only a wave function  for the Universe (U) seen by an external observer (O) could  make sense. But actually it only means that we first divided the whole universe (U') into a sum (U)+(O). So that indeed we can define  $\hat{\rho}^{(red.)}_{U|O}$ and $\hat{\rho}^{(red.)}_{O|U}$.\\
\indent  Now we come to what is the main question or issue of this article and concerns the dilemma discussed in \cite{Brukner,Pienaar} about the `preferred basis problem'. More precisely, we consider an EPR state like 
\begin{eqnarray}
\frac{|\textrm{here}_S\rangle\otimes|\ddot{\smile}_O\rangle+|\textrm{there}_S\rangle_S\otimes|\ddot{\frown}_O\rangle}{\sqrt{2}}\label{eq1}
\end{eqnarray} where $|\textrm{here}_S\rangle,|\textrm{there}_S\rangle $ and $|\ddot{\smile}_O\rangle,|\ddot{\frown}_O\rangle$ are bases for (S) and (O) respectively.
 Eq.~\ref{eq1} can alternatively be written as  
\begin{eqnarray}
\frac{1}{\sqrt{2}}(\frac{|\textrm{here}_S\rangle+|\textrm{there}_S\rangle}{\sqrt{2}})\otimes(\frac{|\ddot{\smile}_O\rangle+|\ddot{\frown}_O\rangle}{\sqrt{2}})\nonumber\\
+\frac{1}{\sqrt{2}}(\frac{|\textrm{here}_S\rangle-|\textrm{there}_S\rangle}{\sqrt{2}})\otimes(\frac{|\ddot{\smile}_O\rangle-|\ddot{\frown}_O\rangle}{\sqrt{2}}).
\label{eq2}\end{eqnarray} As it is well known this leads to the preferred basis problem in quantum measurement theory~\cite{Peres,Zurek}.\\
\indent   Now, once again in RQM for (O) the fundamental quantity is $\hat{\rho}^{(red.)}_{S|O}$ [for (S) this is symmetrically the matrix $\hat{\rho}^{(red.)}_{O|S}=\textrm{Tr}_S[\hat{\rho}_{SO}]$]. Therefore, instead of Eq.~\ref{eq1} or \ref{eq2} we must consider the matrix:
\begin{eqnarray}
\hat{\rho}^{(red.)}_{S|O}=\frac{1}{2}|\textrm{here}_S\rangle\langle \textrm{here}_S|+ \frac{1}{2}|\textrm{there}_S\rangle\langle \textrm{there}_S| \label{eq3}
\end{eqnarray} that is directly equivalent to \cite{Penrose} 
 \begin{eqnarray}
\hat{\rho}^{(red.)}_{S|O}=\frac{1}{2}|\textrm{\underline{here}}_S\rangle\langle \textrm{\underline{here}}_S|+ \frac{1}{2}|\textrm{\underline{there}}_S\rangle\langle \textrm{\underline{there}}_S| \label{eq4}
\end{eqnarray} with \begin{eqnarray}
|\textrm{\underline{here}}_S\rangle=\frac{|\textrm{here}_S\rangle+|\textrm{there}_S\rangle}{\sqrt{2}}\nonumber\\
|\textrm{\underline{there}}_S\rangle=\frac{|\textrm{here}_S\rangle-|\textrm{there}_S\rangle}{\sqrt{2}}.
\end{eqnarray}  The description is thus basis independent. Moreover, the observer (O) doesn't care in RQM about the basis $|\ddot{\smile}_O\rangle,|\ddot{\frown}_O\rangle$ or $\frac{|\ddot{\smile}_O\rangle\pm|\ddot{\frown}_O\rangle}{\sqrt{2}}$ which is not appearing in  $\hat{\rho}^{(red.)}_{S|O}$ and there is no paradox in the mathematical formulation of RQM.\\
\indent Now, this doesn't mean that every thing is clear in the physiczal analysis.   What we showed is that a transparent formalism exists for RQM.   Moreover, the key question in RQM is in my opinion related to the question of the transition from `Potentia' (using Heisenberg reading of Aristotle's dunamis `$\delta\acute{\upsilon}\nu\alpha\mu\iota\varsigma$') to actual\cite{Jaeger}. Indeed, physically RQM is built in analogy with the Copenhagen interpration: A cut is introduced betwen two subsystems (S) and (O) and an asymmetry is introduced.  Indeed,  (O) belongs now to the actualized domain but without an appeal to classicality or `macroscopicality'.   By introducing a trace on (O) variables we  actually mean that (O) has no access, i.e., knowledge, to these variables otherwise (O) could  be a self-observer and we should reintroduce preferred variables.\\
\indent  But now what  are really the status of the (S) subsystem for (O)? Following Heisenberg, we should  consider that  $\hat{\rho}^{(red.)}_{S|O}$ defines a catalog or list of potentialities for possible observations made on (S).  Since $\hat{\rho}^{(red.)}_{S|O}$ is basis independent it describes actually all kinds of possible measurements concerning local observables  $\hat{A}_S$.   However, the question that I found difficult to answer is when and how is realized this transition from this catalog to an actual fact of (S) relatively to (O).  Rovelli, inspired by Heisenberg, discusses in several writing  the key role of interactions generalizing the anthropomorphic notion of measurement.  But the clear definition of an interaction is not obvious. Certainly,  at the formal level in quantum mechanics the introduction of interactions is linked with a coupling hamiltonian $\hat{H}_{SO}$ between the two subsystems. Moreover, in a fully unitary theory  $\hat{H}_{SO}$ is only a mean to transform a factorized quantum state $|\Psi_{S}\rangle\otimes|\Psi_{O}\rangle$ for the composite system into an entangled state $|\Psi_{SO}\rangle$  like the EPR one discussed in Eq.~\ref{eq1}. From the point of view of the formalism the change is only quantitative and not qualitative and therefore something is still missing in my opinion to obtain this actualization wished by Heisenberg and Rovelli.  Of course, Heisenberg could always hide this difficulty in the vague  notion of classicality or `macroscopicality' specific of the Copenhagen interpretation. In the Copenhagen theory we require a collapse \`{a} la von Neumannn~\cite{vonNeumann} that represents this passage from potential to actual and of course it is not clear if this collapse must be physical or epistemic~\footnote{If it is physical we get something like the spontaneaous collapse theory and if it epistemic we obtain either a version of Wigner's interpretation involving the mind~\cite{Wigner} or a version of QBism involving `agents' \cite{Qbism}. }.  This collapse will reintroduce the preferred basis problem now linked to macroscopicality or perhaps defined by some pointer basis using decoherence. But with RQM we want to have a better more universal and democratic framework and therefore something (I believe) must be added in order to avoid the infamous collapse.\\
\indent   I glimpse a natural Bohmian solution to this problem which goes probably beyond RQM:  We must  reintroduce  hidden variables for describing the (S) subsystem from the perspective of (O). That means we must assume that there is no transition from the potential to the actual: Everything is already actualized in (S) but we dont know yet what before the interaction involving $\hat{H}_{SO}$ is switched on. Before a certain time $t_0$ the two subsytems are factorized $|\Psi_{S}\rangle\otimes|\Psi_{O}\rangle$ and we have $\hat{\rho}^{(red.)}_{S|O}=|\Psi_{S}\rangle \langle \Psi_{S}|:=\hat{\rho}_{S}$.  The system (S) is unknown to (O) but something is already actualized or `real' in it. After the interaction starts at time $t>t_0$  (S) and (O) becomes progressively and ideally entangled like in Eq.~\ref{eq1} or \ref{eq2}.  We now obtain a mixture $\hat{\rho}^{(red.)}_{S|O}$  like in Eq.~\ref{eq3} or Eq.~\ref{eq4}.  But since something is actualized from the start this is just a quantitative transformation not a qualitative one and we avoid the mysterious or magical collapse associated with the transition potential$\rightarrow$ actual.   In my option this is the great advantage of introducing an underlying actualized reality  involving hidden variables.   But of course there is a quite high price to pay:  The variables must be hidden (to respect the quantum rules, e.g., Heisenberg's uncertainty principle) and we must reintroduce preferred variables  that were eliminated in the formal description of RQM. In Bohmian mechanics (for fermions) the preferred variables are positions $x_S(t)$  dynamically defined using a guidance equation for the pilot wave \cite{Bohm}:
\begin{eqnarray}
\frac{dx_s(t)}{dt}=F_{SO}(x_S(t),x_O(t),t)
\end{eqnarray}      where $F_{SO}$ is a map that depends on the specific model (for bosonic  field   the beables `$x$' are replaced by field variables). Of course this dynamics is complicated: It requires a preferred time foliation, and we have  nonlocal forces involving beables for (S) and (O) [a similar symmetrical guidance equation exists for $x_O$].   In this approach  it is fundamentally impossible to neglect the role and existence of $x_O(t)$ on the dynamics of $x_S(t)$ unless we have factorized states  $|\Psi_{S}\rangle\otimes|\Psi_{O}\rangle$:  \begin{eqnarray}
\frac{dx_s(t)}{dt}=F_{S}(x_S(t),t).
\end{eqnarray} 
Moreover, we can still use the perspectival or relational   RQM  formalism in Bohmian mechanics but now this is interpreted as a FAPP application. Indeed, if all what is needed is the description of probabilities  or  mean values $\langle \hat{A}_S\rangle$ we can still apply the RQM reduced matrix  like in Eq.~\ref{mean}.  This doesn't contradict the mere formalism of RQM but the high price to pay is to come back with this underlying hidden and absolute reality that was removed from RQM.  A FAPP-RQM has the advantage to unify the different concept of relativist/relational aspects of quantum mechanics   with the absolute ontological reading  of for example de Broglie  Bohm and Einstein.   In that context it is relevant to note that         
if Einstein favored an operational relational perspective when he was founding special relativity (in agreement with the philosophy of Mach), he latter preferred an absolute description   for the general relativity since space and time are now becoming dynamical variables through  the $g_{\mu\nu}(x)$ tensor.  Moreover, Einstein also got interested into the famous `Mach' principle and considered the problem of a mass in rotation in vacuum.    How inertia could appear if the mass is alone in free space?   Progressively  Einstein realized that a completely relational analysis of the problem is not enough: The $g_{\mu\nu}(x)$ tensor has a physical meaning in itself (this Einstein study was also related to his model of a static Universe  before he introduced the famous cosmological constant).  Of course  the relational reading is not eliminated. Time is defined by physical clocks  and clocks are in turn affected by mass and the tensor $g_{\mu\nu}(x)$.  The two relational and absolute perspectives are thus in my opinion inseparable, i.e.,  entangled, both in general relativity and in quantum mechanics.  We can not have relations without relata  and reciprocally  relata without relations, i.e., pure isolated systems, have no physical meaning.  This   remarkable symmetry is perhaps the key feature that will one day allow us to unify the two theories and therefore RQM is for me extremelly valuable as a  methodological guide.      

\section{Acknowledgments} 
I wish to thank Claudio Calosi for inviting me to participate to this special issue concerning RQM  and   also Carlo Rovelli for very interesting discussions concerning the physical meaning of RQM.  I emphasize that the view presented here concerning RQM is only mine and is not necessarily shared by Calosi or Rovelli.\\
\indent Note added in proof: After this work was completed a critical analysis of RQM was  discussed in \cite{Zukowski2022} and commented by  the present author in \cite{DrezetZukowski2022} (see also the reply \cite{Zukowskireply2022}). I believe that the present analysis of RQM actually constitutes a reply to  \cite{Zukowski2022}.


\begin{thebibliography}{}
\bibitem{Rovelli1996}
Rovelli, C.: Relational quantum mechanics. Int. J. Phys. \textbf{35}, 1637 (1996). 
\bibitem{RovelliBook2021}
Rovelli, C.: Helgoland: Making sense of the quantum revolution, Riverhead Books, New York (2021). 
\bibitem{RovelliFP}
Di Biagio, A., Rovelli, C.: Stable Facts, Relative Facts. Found. Phys. \textbf{51}, 30 (2021).
\bibitem{Bohr}
Bohr, N.: Can quantum-mechanical description of physical reality be considered complete? Phys. Rev. \textbf{48}, 696 (1935).
\bibitem{Bell}
Bell, J.S.: Speakable and Unspeakable in Quantum Mechanics,  2nd edn. Chap. 23. Cambridge University Press, Cambridge (2004).
\bibitem{Bitbol}
Bitbol, M.: De l'int\'erieur du monde, Flammarion, Paris (2010). 
\bibitem{Laudisa}
Laudisa, F., Rovelli, C.: Relational quantum mechanics. In: Zalta, E.N. (ed.) The Stanford Encyclopedia of Philosophy, winter 2019 edn. Metaphysics Research Laboratory, Stanford University, Stanford (2019). https://plato.stanford.edu/archives/win2019/entries/qm-relational/
\bibitem{Peres}
Peres, A.: When is a quantum measurement? Am. J. Phys.~\textbf{54}, 688 (1988). 
\bibitem{Brukner}
Brukner, C.: Qubits are not observers – a no-go theorem (2021). https://arxiv.org/abs/2107.03513v1 [quant-ph]
\bibitem{Pienaar}
Pienaar, J.: A Quintet of Quandaries: Five No-Go Theorems for Relational Quantum Mechanics. Found. Phys. \textbf{51}, 97 (2021). 
\bibitem{Rovellian}
Di Biagio, A., Rovelli, C.: Relational Quantum Mechanics is about Facts, not States: A reply to Pienaar and Brukner (2021). https://arxiv.org/abs/2110.03610 [quant-ph]
\bibitem{Simon}
Simon, C.: Conscious observers clarify many worlds. arXiv:0908.0322v1. (2009). 
\bibitem{Zurek}
Zurek, W.H.: Pointer basis of quantum apparatus: Into what mixture does the wave packet collapse? Phys. Rev. D \textbf{24}, 1516 (1981). 
\bibitem{Penrose}
Penrose, R.: The large, the small and the human mind. Cambridge University Press, Cambridge (1997).
\bibitem{Jaeger}
Jaeger, G.: Quantum potentiality revisited, Phil. Trans. R. Soc. A \textbf{375}, 20160390 (2017).
\bibitem{vonNeumann}
von Neumann, J.:   Mathematische Grundlagen der Quantenmechanik. Berlin, Springer (1932). (English translation Mathematical Foundations of Quantum
Mechanics, Princeton, Princeton University Press (1955)).
\bibitem{Wigner}
Wigner, E. P.:  Remarks on the mind-body question. In The scientific Speculates; I. J. Good Ed. Heinemann, London (1961).
\bibitem{Qbism}
Fuchs, C.A., Mermin, N. D., Schack, R.: An introduction to QBism with an application to the locality of quantum mechanics. Am. J. Phys. \textbf{82}, 749 (2014).
\bibitem{Bohm} Bohm., D and Hiley, B.J.: The undivided Universe. Routledge, London, (1993).
\bibitem{Zukowski2022}
Lawrence, J., Markiewicz, M., Zukowski, M.: Relative facts do not exist. Relational Quantum Mechanics is Incompatible with Quantum Mechanics. arXiv:2208.11793 (2022).
\bibitem{DrezetZukowski2022}
Drezet, A.: (Once more) In defense of Relational Quantum Mechanics: A note on `Relative facts do not exist. Relational quantum mechanics is incompatible with quantum mechanics'. arXiv:2209.01237 (2022).
\bibitem{Zukowskireply2022}
Lawrence, J., Markiewicz, M., Zukowski, M.: Relative facts do not exist. Relational Quantum Mechanics is Incompatible with Quantum Mechanics.  Response to the critique by Aur\'{e}lien Drezet. arXiv:2210.09025 (2022).
\end{thebibliography}
\end{document}